\documentclass[ 
 preprint,
nofootinbib,
 amsmath,amssymb,
 aps]{revtex4-2}

\usepackage{comment}
\usepackage{graphicx}
\usepackage{dcolumn}
\usepackage{bm}
\usepackage{hyperref}
\hypersetup{colorlinks=true, linkcolor=magenta, filecolor=cyan, urlcolor=blue}

\usepackage{xcolor}
\def\Mpl{M_{\rm P}}

\begin{document}

\preprint{IPMU26-0010}

\title{Branching Universes}

\author{Anamaria Hell}
\email{anamaria.hell@ipmu.jp}
\affiliation{Kavli IPMU (WPI), UTIAS, The University of Tokyo,\\
and Center for Data-Driven Discovery,\\
5-1-5 Kashiwanoha, Kashiwa, Chiba 277-8583, Japan}

\author{Tatsuya Daniel}
\email{tdaniel@physics.mcgill.ca}
\affiliation{Trottier Space Institute, Department of Physics, McGill University, Montreal, QC H3A 2T8, Canada}

\date{September 11, 2026}

\begin{abstract}

We propose the idea that our Universe is a realization among different possible branches, which can be observationally tested through the modified dispersion relation of the gravitational waves. We achieve this through a framework of spatially constrained vector fields.  We show that the simplest realizations of such theories in flat and cosmological spacetimes do not introduce new propagating modes, but they give rise to tensor perturbations that differ from those of standard general relativity. As a result, they avoid the strong coupling issues or ghost modes that otherwise easily appear in vector-field models. We further show that such theories admit stealth black hole solutions, and we recover weak gravitational potentials, thus passing the solar system experiments. Finally, we discuss the implications of such theories and propose further generalizations.

\end{abstract}

\maketitle

\newpage

\tableofcontents

\section{\label{sec:level1}Introduction }
General relativity (GR) is one of the simplest and most elegant theories of gravity that describes our Universe. Yet, it stands challenged, with foundational theoretical puzzles and through the recent tests of $\Lambda$CDM, while at the same time overcoming the solar system experiments with nearly perfect success.

One of the key predictions of GR is that gravitational waves (GWs) propagate at the speed of light \cite{Misner:1973prb,Carroll:2004st}. Proving this dispersion relation is even more exciting through LIGO-Virgo-KAGRA (LVK) and data from pulsar timing arrays (PTAs), as well as upcoming observations from the Laser Interferometer Space Antenna (LISA) \cite{LISA:2022yao} and the Square Kilometre Array (SKA) \cite{Dewdney2009}. However, with an increasing precision, it may also happen the speed is modified \cite{LIGOScientific:2017zic}. In this work, we propose that depending on this value, our Universe might be just one of the realizations among many possible branches.  

In light of GR challenges, modified theories of gravity have become an appealing alternative that might underline the gravitational force. 
Such theories often give rise to new propagating modes -- which can manifest themselves through additional GW polarizations, such as in massive gravity \cite{Fierz:1939ix, Chamseddine:2010ub, deRham:2010kj} -- or they introduce additional matter fields, as, for example, in scalar-tensor theories \cite{Brans:1961sx, Horndeski:1974wa, Deffayet:2011gz, Kobayashi:2011nu, Langlois:2015cwa, Crisostomi:2016czh, BenAchour:2016fzp}, mimetic gravity \cite{Chamseddine:2013kea}, or the constrained scalar field framework \cite{Hell:2025lgn}.

In contrast, one can also modify gravity such that no additional modes propagate. This possibility was realized in cuscuton gravity \cite{Afshordi:2006ad}, which belongs to the class of theories in which Lorentz invariance is violated \cite{Afshordi:2009tt, Bhattacharyya:2016mah, Gomes:2017tzd}. Modifying gravity without additional propagating modes was also realized in extensions of cuscaton gravity \cite{ Iyonaga:2018vnu}, as well as more generally in minimally-modified theories of gravity \cite{Lin:2017oow, Mukohyama:2019unx, Carballo-Rubio:2018czn, Aoki_2019, DeFelice:2020ecp, DeFelice:2020eju}.

Following this ideology, one might naturally ask: \textit{Can we construct a 4-dimensional diffeomorphism-invariant theory of gravity that is easily distinguishable from GR even in the simplest cases, such as the de Sitter Universe and in flat spacetime, but which only propagates two gravitational modes?}  In this work, we show that this is possible to achieve with the help of specially constrained vector field theories that are non-minimally coupled to gravity.

Modifications of gravity based on vector fields have so far proven to be of significant cosmological relevance. The standard theories show great promise to underline the origin for primordial magnetic fields \cite{Turner:1987bw, Demozzi:2009fu}, dark matter \cite{Essig:2013lka, Graham:2015rva, Kolb:2020fwh}, as well as the early- and late-time acceleration of the Universe \cite{Ford:1989me, Koivisto:2008xf, Golovnev:2008cf, Alexander:2014uza, DeFelice:2016yws}. They also give rise to birefringence in parity-violating theories \cite{Preuss2004,Daniel_2024,Alexander:2025wnj}. 

However, models based on vector fields often face difficulties. In the presence of Ricci tensor couplings, theories based on the Proca field give rise to the strong coupling of tensor perturbations in flat spacetime, at energies below the Planck mass\footnote{The same issue is present even in the Generalized Proca theory \cite{Heisenberg:2014rta} due to the presence of the coupling with the Ricci tensor, and also occurs for the Kalb-Ramond field with non-minimal coupling to gravity \cite{Hell:2026mle}.} \cite{Hell:2024xbv}. This implies that regardless of how small the photon mass is, the GWs would be strongly coupled, posing a serious theoretical inconsistency. The coupling with the Ricci tensor also gives rise to a runaway behavior if the vector field is taken as a spectator field in a homogeneous and isotropic background \cite{Capanelli:2024pzd}. 

One possibility to avoid both of these issues is to instead formulate the theory in the disformal frame \cite{Hell:2024xbv}. However, in the presence of non-minimal coupling to gravity, the vector field can also contribute to the cosmological expansion. While in this case the coupling with the Einstein tensor describes a single scalar mode \cite{DeFelice:2016yws}, for a general non-minimal coupling with Ricci scalar and Ricci tensor, the theory describes an additional scalar perturbation whose speed of propagation  vanishes, thus indicating another type of breakdown of the perturbation theory \cite{DeFelice:2025ykh}. The vanishing speed of propagation also appears for one of the modes if one replaces the Maxwell kinetic term with a dual of the 3-form, and studies the theory in the anisotropic background \cite{DeFelice:2025khe}.    

These issues naturally motivate a search for an alternative theory of vector fields that could give rise to interesting descriptions of cosmological expansion or black holes, while at the same time avoiding the previous problems. In this study, we will present a realization of this goal by formulating one of the simplest vector-field theories, based on the recently proposed framework of \textit{constrained gravity} \cite{Hell:2025lgn}. In this framework, gravity is modified through fields with no kinetic terms but with non-minimal couplings to gravity, which give rise to different branches along which the Universe can expand. While its simplest formulation based on the scalar field is very interesting on its own, we will show that a generalization of this framework to the vector fields gives rise to a theory that is easily testable and intriguing, while having no kinetic terms for the vector field.

In particular, by focusing on the de Sitter spacetime, we will show that the simplest realization of our theory can give rise to two branches, with vanishing and non-vanishing values of the vector field that describe only GWs, with no extra modes propagating. Moreover, we will find that while the first branch recovers standard GR and gives rise to unit speed of propagation for the GWs, if the GW propagation speed is ever so slightly shifted, our Universe will be governed by the second branch, which has non-vanishing values of the vector field. In addition, we will show that this branch can give rise to interesting cosmological coupling with external matter, which directly builds on the framework of the constrained scalar. Furthermore, the theory admits stealth black hole solutions, which gives rise to even more branches.

The paper is organized as follows. First, in Section~\ref{sec:theory} we will present the general theory, and discuss its key properties in de Sitter and flat spacetimes. Then, in Section~\ref{sec:matter}, we will lay out the key equations for the cosmological expansion, and show the gravitational potentials and black-hole solutions. In Section~\ref{sec:implications}, we will consider immediate implications of this framework, before summarizing our results in Section~\ref{sec:discussion}.

\section{Branches of vector fields }\label{sec:theory}

Most of the standard approaches to vector fields involve theories that contain standard kinetic terms\footnote{There are theories with exceptions outside of the Maxwell and 3-form kinetic terms, such as the vector formulation of unimodular gravity \cite{Henneaux:1989zc}, which have no explicit kinetic terms of the vector fields.}. In this work, we will consider another possibility, that the vector fields are sourced purely by gravity, with the action of the following form: 
\begin{equation}\label{pureVF}
    S=\int d^4x\sqrt{-g}\left[\frac{\Mpl^2}{2}\left(R-2\Lambda\right)-m^2A_{\mu}A^{\mu}+\frac{1}{2}\xi RA_{\mu}A^{\mu}-\xi R^{\mu\nu}A_{\mu}A_{\nu}\right],
\end{equation}
where $\Lambda$ is the cosmological constant and $\xi$ is the coupling. 

At first glance, the above action can be counter-intuitive: when thinking of vector fields, or even fields in general, one adds the kinetic terms as part of the standard procedure. However, here, we instead omit them, and consider the spacetime evolution driven by these constrained fields. Due to the non-minimal coupling, we will see that such a reduction can give rise to interesting non-trivial background dynamics: among possible solutions, it will include two de Sitter Universes that differ in the values of the curvature.

One should note that the above coupling takes a very particular form, with the Ricci tensor and Ricci scalar forming the Einstein-tensor coupling. This is necessary to keep gravity minimally modified, and not introduce additional propagating modes at least for the cosmological backgrounds. Otherwise, if the couplings with the Ricci scalar and Ricci tensor take a more arbitrary form, the theory will propagate additional scalar degree of freedoms, as we will show in Appendix \ref{gendof}.   

\noindent
By varying the action \eqref{pureVF} with the vector field, we find the following equation of motion: 
\begin{equation}
    G^{\mu\nu}A_{\nu}+m^2A^{\mu}=0,
\end{equation}
where $G_{\mu\nu}$ is the Einstein tensor. By varying the action with respect to the metric we find:
\begin{equation}
    \begin{split}
        &\frac{\Mpl^2}{2}\left(G_{\mu\nu}+g_{\mu\nu}\Lambda\right)-\left(m^2-\frac{\xi}{2}R\right)A_{\mu}A_{\nu}+\frac{1}{2}g_{\mu\nu}\left(m^2A_{\gamma}A^{\gamma}+\xi G^{\gamma\delta}A_{\gamma}A_{\delta}\right)\\&
        +\frac{\xi}{2}\left(\nabla_{\gamma}\nabla^{\gamma}f_{\mu\nu}-\nabla_{\gamma}\nabla^{\mu}f_{\nu}^{\gamma}-\nabla_{\gamma}\nabla^{\nu}f_{\mu}^{\gamma}+g_{\mu\nu}\nabla_{\gamma}\nabla_{\delta}f^{\gamma\delta}\right)=0,
    \end{split}
\end{equation}
where 
$f^{\mu\nu}=g^{\mu\nu}A_{\gamma}A^{\gamma}-A^{\mu}A^{\nu}$.  Let us now analyze the simplest cases: the de Sitter branches and the flat spacetime.

\subsection{The two de Sitter branches}\label{subsec:dSa0}

As a first step, let us analyze the free theory in (\ref{pureVF}) assuming a homogeneous and isotropic Universe. For this, we will focus on the background solutions, around which we will perturb, following the recipe to finding the degrees of freedom \cite{Hell:2026blj}.  We will assume that the metric and the vector field take the following background values: 
\begin{equation}\label{ansatzdS}
   \begin{split}
    &ds^2=-N^2dt^2+a^2\delta_{ij}dx^idx^j\qquad \text{and}\qquad   A_{\mu}=(N A_0(t),0,0,0),
     \end{split}
\end{equation}
where $N$ is the lapse. By varying with respect to $N$, and setting it to unity, we find the constraint equation: 
\begin{equation}
    \begin{split}
   \frac{A_{0}^{2} m^{2}}{2}-\Lambda \,\Mpl^{2}+\frac{3 A_{0}^{2} \dot{a}^{2} \xi}{a^{2}}+\frac{3 \Mpl^{2} \dot{a}^{2}}{a^{2}}=0.
    \end{split}
\end{equation}
Next, by varying with respect to the scale factor, we find the acceleration equation: 
\begin{equation}
    \begin{split}
        A_{0}^{2} m^{2} a^{2}-2 \Lambda \,\Mpl^{2} a^{2}+8 \dot{A}_0 A_{0} \dot{a} \xi  a +2 A_{0}^{2} \dot{a}^{2} \xi +4 A_{0}^{2} \ddot{a} \xi  a +2 \Mpl^{2} \dot{a}^{2}+4 \Mpl^{2} \ddot{a} a
=0,
    \end{split}
\end{equation}
while the constraint for $A_0$ yields: 
\begin{equation}
 A_{0} \left(m^{2}-\frac{6 \xi  \dot{a}^{2}}{a^{2}}\right)=0.
\end{equation}
We can notice that the last relation gives rise to two branches for the background. In particular, if we assume that $A_0\neq0$, we find an equation that sets the Hubble parameter to a constant: 
\begin{equation}\label{Hubble1}
   H_{dS_1}^2=\frac{m^2}{6\xi}. 
\end{equation}
By substituting this solution into the constraint equation, one finds that the temporal component of the vector field also takes a constant value: 
\begin{equation}\label{dSA0sol}
    A_0^2=\frac{\Mpl^{2} \left(2 \Lambda \xi -m^{2}\right)}{2 \xi  \,m^{2}}, 
\end{equation}
while the acceleration equation becomes identically satisfied. The above relation gives us a bound on the coupling. To ensure that $A_0^2$ is positive, we must have
\begin{equation}\label{A0condition}
    2 \Lambda \xi -m^{2}>0\qquad \text{and}\qquad \xi m^2>0,
\end{equation}
or both inequalities smaller than zero. 

In addition to this de Sitter branch, where the Hubble parameter is set by (\ref{Hubble1}), the theory admits another de Sitter branch for vanishing values of the temporal component. In this case, the Hubble parameter is set by the constraint equation to be: 
\begin{equation}\label{Hubble2}
    H_{dS_2}^2=\frac{\Lambda}{3}. 
\end{equation}

After perturbing around the background, with the procedure detailed in Appendix \ref{appA}, we find the following result: both scalar and vector modes are not contained in the theory, while the tensor modes propagate, with the Lagrangian density given by:
\begin{equation}\label{LagrtendS}
    \mathcal{L}_{TdS_1}=\frac{a^3(\Mpl^2+\xi\Theta)}{8}\left(\dot{h}^T_{ij}\dot{h}^T_{ij} -\frac{1}{a^2}\frac{(\Mpl^2-\xi\Theta)}{(\Mpl^2+\xi\Theta)}h_{ij,k}^Th_{ij,k}^T\right),
\end{equation}
where $\Theta=A_0^2$ and $_{,k}=\frac{\partial}{\partial x^k}$. Their speed of propagation at high momentum is given by: 
\begin{equation}\label{cTUniversal}
     c_{{TdS_1}}^2=\frac{(\Mpl^2-\xi\Theta)}{(\Mpl^2+\xi\Theta)}. 
\end{equation}
This means that for the GWs to be well-behaved, we require: 
\begin{equation}\label{boundT}
    \Mpl^2>\theta\xi>-\Mpl^2.
\end{equation}
In terms of the solution (\ref{dSA0sol}) for $A_0$, and taking into account (\ref{A0condition}), this can be interpreted as a bound on $m^2$:
\begin{equation}
    2\Lambda\xi>m^2>\frac{2\Lambda\xi}{3} .\label{eq:m2-ghost-bound}
\end{equation}

Similar to the nonzero $A_0$ branch (determined by the equations (\ref{Hubble1}) and 
(\ref{dSA0sol})), in the second branch, where $A_0=0$ and the Hubble parameter is given by (\ref{Hubble2}), the scalar and vector modes also do not propagate. On the other hand, the tensor modes do propagate, with the Lagrangian density given by the standard expression:
\begin{equation}
    \mathcal{L}_{TdS_2}=\frac{\Mpl^2a^3}{8}\left(\dot{h}_{ij}^T\dot{h}_{ij}^T-\frac{1}{a^2}h_{ij,k}^Th_{ij,k}^T\right),
\end{equation}
which gives rise to unit speed of propagation for the GWs:
\begin{equation}
    c_{TdS_2}^2=1.
\end{equation}

\subsection{\label{subsec:flat}Branches of flat spacetime}
A natural question is to ask if such a scenario allows for a realization of the flat spacetime. In this case, the constraint equation yields
\begin{equation}
    m^2A_0=0, 
\end{equation}
and thus we have two possibilities: either we set $A_0=0$, or $m=0$. By choosing vanishing mass, we find that $\Lambda=0$. However, none of the background equations of motion determine $A_0$, meaning that it has now become completely arbitrary. To investigate its influence on the propagation of GWs, we expand the action again up to second order in perturbations.  After integrating by parts, we find: 
\begin{equation}\label{LgravFlat1}
    \mathcal{L}_{Tflat1}=\frac{(\Mpl^2+\xi\Theta)}{8}\left(\dot{h}^T_{ij}\dot{h}^T_{ij} -\frac{(\Mpl^2-\xi\Theta)}{(\Mpl^2+\xi\Theta)}h_{ij,k}^Th_{ij,k}^T\right).
\end{equation}
Therefore, we find that their speed of propagation matches the de Sitter case (\ref{cTUniversal}), with the bound determined by (\ref{boundT}). However, we should note that $A_0$ is not a gauge mode. While it takes an arbitrary value on the level of the background, it can nevertheless be measured by determining the speed of propagation, indicating that it is a physical variable. This suggests that we could think of $A_0$ as a sort of vacuum value. In this way, one might be able to observationally test which branch we are in by measuring the GW propagation speed. The LVK constraints may tell us if/what non-zero values of $A_0$ could be allowed. We discuss this point more in Sec.~\ref{sec:lvk-observations}.

One should note that if we instead set $A_0=0$ on the background, then the remaining equations allow for an arbitrary $m$, while the cosmological constant has to be set to zero to recover the Minkowski spacetime. In this case, however, the GWs will satisfy: 
\begin{equation}
    \mathcal{L}_{Tflat2}=\frac{\Mpl^2}{8}\left(\dot{h}_{ij}^T\dot{h}_{ij}^T-h_{ij,k}^Th_{ij,k}^T\right),
\end{equation}
which can also be smoothly obtained from (\ref{LgravFlat1}) by setting $A_0=0$, and the mass will not play a role in this case. 

\section{ The key solutions \label{sec:matter} }

\subsection{Cosmological evolution }

Let us now study how to couple matter to our theory. As we will discuss in Sec.~\ref{subsec:framework}, due to the vector fields, this coupling can take quite a general form. In this subsection, we will focus only on the simplest cases, with the action given by:
\begin{equation}\label{s1caseCosmo}
  S_1=\int d^4x\sqrt{-g}\left[\frac{\Mpl^2}{2}\left(R-2\Lambda\right)-m^2A_{\mu}A^{\mu}+\frac{1}{2}\xi RA_{\mu}A^{\mu}-\xi R^{\mu\nu}A_{\mu}A_{\nu}+ f(A_{\mu}A^{\mu})p\right],
\end{equation}
where $p$ is the pressure and $f$ is an arbitrary function of $A_{\mu}A^{\mu}$. If $f=1$, we can notice that the external matter is minimally coupled to gravity and the vector field, while otherwise it becomes non-minimally coupled\footnote{The (non-perturbative) Einstein frame does not exist in (\ref{s1caseCosmo}) due to the coupling with the Ricci tensor. However, the theory does admit the minimal frame, where matter is minimally coupled to both the external field and to gravity \cite{Hell:2025lgn}.}. In the following, we will first study the minimally coupled case, and then with $f=\beta A_{\mu}A^{\mu}$, where $\beta$ is the coupling constant.

\subsubsection{The standard matter coupling}

Let us first consider the case with $f=1$. In this case, we find that the constraint equations, obtained by varying the action with respect to the lapse $N$, and then setting $N$ to unity, is given by:
\begin{equation}
    H^2=\frac{-m^{2} \Theta+2 \Mpl^{2} \Lambda +2   \varepsilon}{6\left( \Theta \xi + \Mpl^{2}\right)},
\end{equation}
where $\varepsilon$ is the energy density of matter, which satisfies:
\begin{equation}
    \dot{\varepsilon}=-3H(\varepsilon+p).
\end{equation}
The acceleration equation, found by varying the action with respect to the scale factor, is given by:
\begin{equation}
    \frac{\ddot{a}}{a}=-\frac{ 6  H \dot{\Theta} \xi +m^{2} \Theta-2 \Mpl^{2} \Lambda +3   p +  \varepsilon}{6\left( \Theta \xi + \Mpl^{2}\right)}.
\end{equation}
Finally, the temporal component gives rise to the following equation:
\begin{equation}\label{A0constraintstan}
    A_0 \left(m^{2}-6 \xi H^{2}\right)=0,
\end{equation}
which gives rise to two branches for $A_0$.

If $A_0\neq 0$, the background evolution corresponds to the de Sitter Universe due to (\ref{A0constraintstan}), in which $A_0$ can now also evolve in time in contrast to Sec.~\ref{subsec:dSa0}. This will affect the speed of propagation of the GWs, whose Lagrangian density and the speed of propagation match with the expressions (\ref{LagrtendS}) and (\ref{cTUniversal}). 

In contrast to the $A_0\neq 0$ branch, one can easily see that setting $A_0=0$ will recover the standard cosmology with unit speed of propagation for the GWs. 

\subsubsection{The non-minimal coupling}

Assuming that $f=\beta A_{\mu}A^{\mu}$ leads to the case in which external matter is non-minimally coupled to the vector field. Here, we will study this case, assuming that the matter is of the form of a perfect fluid. In this case, the constraint equation is given by: 
\begin{equation}\label{constraintNmin}
    H^2=\frac{-2 \beta  \Theta \varepsilon-m^{2} \Theta+2 \Lambda \,\Mpl^{2}}{6 \Theta \xi +6 \Mpl^{2}},
\end{equation}
while the acceleration equation generalizes to:
\begin{equation}
    \ddot{a}=-\frac{a \left(\left(m^{2}-\beta  \varepsilon-3 \beta  p\right) \Theta+6 \dot{\Theta} H  \xi -2 \Lambda \,\Mpl^{2}\right)}{6 \Theta \xi +6 \Mpl^{2}}.
\end{equation}
The equation of motion for the $A_0$ component is given by:
\begin{equation}
    \sqrt{\Theta} \left(-6 \xi  H^{2}-2 \beta  p+m^{2}\right)=0.
\end{equation}
Therefore, similar to the previous cases, depending on $A_0$ we have two branches. Let us first consider the $A_0\neq 0$ case, which results in $ -6 \xi  H^{2}-2 \beta  p+m^{2}=0$. 
Due to the non-minimal coupling, the conservation equation for the perfect fluid now acquires an extra term:
\begin{equation}
    \dot{\varepsilon}=-\left(3H+\frac{\dot{\Theta}}{\Theta}\right)\left(\varepsilon+p\right).
\end{equation}
We can notice that the above equations differ significantly from the standard Friedmann equations. The main change is the time-evolving $A_0$ component, which introduces an additional contribution even at the level of the conservation equation. 

However, we can see that despite this, there is a special case in which the equations resemble those governing the standard Big Bang cosmology. In particular, by setting $m=0$, $\Lambda=0$, assuming that the temporal component is constant ($\dot{\Theta}=0$), and identifying
\begin{equation}
  p=\frac{\Theta\xi}{M^2}\varepsilon\qquad \text{and}\qquad \frac{\beta\Theta}{6M^2}=-\frac{4\pi G}{3}, 
\end{equation}
where $M^2=\Theta\xi+\Mpl^2$, the above equations take the well-known form of the Friedmann equations and continuity equation. So, if we assume standard cosmology like we have above with the non-minimal coupling, it would suggest that by measuring $G$ through the gravitational potential, we could also constrain which branch of $A_0$ we are in, although this might change depending on the cosmological model.

Even though we can recover the standard cosmological evolution, this does not mean that our theory is indistinguishable from standard GR. In particular, it allows for a more general set of solutions since $A_0$ can be time-evolving. Moreover, even the special case of standard background cosmological evolution can still differ due to the metric perturbations. In particular, by varying the metric with respect to the tensor modes around the above background, we again recover the Lagrangian density and GW speed of propagation that matches with the previous relations in (\ref{LagrtendS}) and (\ref{cTUniversal}). Therefore, we also find here that the GWs propagate at a speed different from the speed of light, making this possibility easily testable. 

\subsection{Gravitational potentials }\label{sec:gravitational-potentials}

In order to study the gravitational potentials, and thus verify if our theory admits the solar system tests, we will start our analysis from the spherically symmetric ansatz: 
\begin{equation}
    ds^2=-e^{2\alpha(r)}dt^2+e^{2\beta(r)}dr^2+r^2d\theta^2+r^2\sin^2{\theta}d\varphi^2,\qquad A_{\mu}=(A_0(r),0,0,0), 
\end{equation}
and set $m=0$ and $\Lambda=0$. Then, by substituting this into the action, and by varying the action with respect to $\alpha$ and $\beta$, we find respectively:
\begin{equation}
    \left(\Mpl^{2} {\mathrm e}^{2 \alpha}+A_0^{2} \xi \right) \left({\mathrm e}^{2 \beta}+2 \beta' r -1\right)=0,
\end{equation}
and
\begin{equation}
    \left(\left(\alpha' r -\frac{1}{2}\right) A_0-2 r A_0'\right) A_0 \xi  \,{\mathrm e}^{-2 \beta-2 \alpha}+\Mpl^{2} \left(\alpha' r +\frac{1}{2}\right) {\mathrm e}^{-2 \beta}+\frac{\xi  \,{\mathrm e}^{-2 \alpha} A_0^{2}}{2}-\frac{\Mpl^{2}}{2}
=0,
\end{equation}
while varying with respect to $A_0(r)$ yields:
\begin{equation}
    -2 \xi  A_0 \left({\mathrm e}^{2 \beta}+2 \beta' r -1\right)=0,
\end{equation}
with primes denoting the derivative with respect to $r$. One can easily check that, assuming that $A_0\neq 0$, the theory admits the Schwarzschild solution:
\begin{equation}\label{schbh}
     ds^2=-\left(1-\frac{r_g}{r}\right)dt^2+\left(1-\frac{r_g}{r}\right)^{-1}dr^2+r^2d\theta^2+r^2\sin^2{\theta}d\varphi^2,
\end{equation}
with

\begin{equation}
    A_0(r)=C\sqrt{\frac{r-r_g}{r r_g}},
\end{equation}
where $r_g=2M$ is the Schwarzschild radius, and $C$ is the constant of integration. Note that in this case $r$ cannot go to zero, as it would make the square root imaginary. However, the theory admits the weak gravitational potentials, and thus by expanding the above solution we can easily recover matching with GR. In particular, at asymptotic infinity, given by the limit when $r\to\infty$, we smoothly recover flat spacetime, with constant values of $A_0$ at leading order. 

\subsection{Stealth black holes}\label{stealthBHsec}

Notably, our theory also admits stealth black holes -- a special type of black hole that cannot be distinguished from the standard GR solution at background order, but which admits non-trivial values for the background fields (see, for example \cite{Babichev:2023psy} and references therein). In our case, such a solution can be obtained by setting $m=0$ and $\Lambda=0$, and substituting the following ansatz into the action,
\begin{equation}
    ds^2=-e^{2\alpha(r)}dt^2+e^{2\beta(r)}dr^2+r^2d\theta^2+r^2\sin^2{\theta}d\varphi^2,\qquad A_{\mu}=(A_0(r),A_1(r),0,0). 
\end{equation}
This leads us to the following system of equations: 
\begin{equation}\label{StealthBHeqs}
    \begin{split}
        0=&{\mathrm e}^{-2 \alpha} A_0^{2} \xi +{\mathrm e}^{-4 \beta} \xi  \left(-6 \beta' r A_1^{2}+4 A_1' A_1 r +A_1^{2}\right)+\Mpl^{2}+{\mathrm e}^{-2 \beta} \left(2 \Mpl^{2} \beta' r +A_1^{2} \xi -\Mpl^{2}\right)\\&+{\mathrm e}^{-2 \beta-2 \alpha} A_0^{2} \xi  \left(2 \beta' r -1\right),
        \\0=& {\mathrm e}^{-2 \alpha} A_0^{2} \xi -3 \,{\mathrm e}^{-4 \beta} A_1^{2} \xi  \left(2 \alpha' r +1\right)-\Mpl^{2}-{\mathrm e}^{-2 \beta} \left(-2 \alpha' \Mpl^{2} r -A_1^{2} \xi -\Mpl^{2}\right)\\&-{\mathrm e}^{-2 \beta-2 \alpha} \xi  \left(-2 \alpha' r A_0^{2}+4 A_0' A_0 r +A_0^{2}\right),\\
        0=&-2 \xi  A_0 \left(2 \beta' r +{\mathrm e}^{2 \beta}-1\right),\\
        0=& -A_1^{2}  \xi^{2} \left(2 \alpha' r -{\mathrm e}^{2 \beta}+1\right). 
    \end{split}
\end{equation}
 Notably, these equations admit the Schwarzschild black hole described by the metric (\ref{schbh}) with:
\begin{equation}
    A_1=\frac{c_{1} \sqrt{r}}{r r_g -1},   \qquad \text{and}\qquad A_0^2=\frac{r_g^{3} c_{2} r -c_{2} r_g^{2}+c_{1}^{2}}{r \,r_g^{2}},
\end{equation}
where $c_1$ and $c_2$ are constants of integration. Note here that since $A_1 = 0$ is the special case, which can be found from the constraint in \eqref{StealthBHeqs}, we have different branches corresponding to $A_1 = 0$ and $A_1 \neq 0$. Therefore, the above case has multiple branches, since now both $A_0$ and $A_1$ can have vanishing and non-vanishing values.

\section{Implications} \label{sec:implications}

\subsection{LVK observations} \label{sec:lvk-observations}
The deviation of the GW propagation speed from the speed of light has been constrained to be at most about one part in $10^{15}$ \cite{LIGOScientific:2017zic}, a bound which continues to be improved by detections from the LVK collaboration. While this observation immediately ruled out many modified theories of gravity, we demonstrate here that we can use this constraint to observationally test which branch(es) our Universe is currently in.

As an example, let us take (\ref{cTUniversal}) as a form for the modified dispersion relation. 
 We can write the propagation speed constraint from \cite{LIGOScientific:2017zic} as $ |c_T - 1| \lesssim 10^{-15}$, which when combined with (\ref{cTUniversal}) yields
$|\Theta\xi|\lesssim 10^{-15}\Mpl$. 
Therefore, an observation of a nonzero $\Theta\xi$ would suggest the presence of healthy branches. On the other hand, as soon as the deviation of the GW speed from $c$ vanishes, it implies that the Universe has realized the other branch of $\Theta = 0$.

From LVK, one can constrain $\xi\Theta$, but because the dispersion relation (\ref{cTUniversal}) appears to be universal, as it applies to different spacetimes, one may later be able to use the LVK constraint on $\xi\Theta$ to also constrain cosmology.

\subsection{Gravitational counterparts of fast radio bursts }

Strongly-lensed FRBs experience a phenomenon known as \textit{gravitational slip}, where (in our convention) the gravitational potential $\psi$ becomes unequal to the time-time component of the metric perturbation in Newtonian gauge $\phi$.  
The parameter $\gamma$ describing the amount of slip is defined as $\gamma \equiv \psi/\phi$, with $\gamma = 1$ corresponding to GR. It has been shown that FRB time-delay measurements can yield constraints as tight as 
\begin{align}
    |\gamma_{\text{PN}} - 1| \lesssim 0.04~\times~(\lambda/100~\text{kpc})~\times~[N/10]^{-1/2}, \label{eq:grav-slip}
\end{align}
where $\gamma_{\text{PN}}$ is the post-Newtonian gravitational slip parameter, $\lambda$ is the supergalactic screening scale, and $N$ is the number of systems \cite{Adi2021}. 

As we showed in Sec.~\ref{sec:gravitational-potentials}, even though $A_0$ can modify the gravitational potentials, it does not change the ratio $\gamma$. Therefore, this branch admits black hole solutions that are observationally allowed by the gravitational FRB constraints. In turn, (\ref{eq:grav-slip}) may be used to test GR and our theory, since they both pass the solar system tests. Furthermore, we may be able to use (\ref{eq:grav-slip}) to test whether other general black hole solutions corresponding to different nonzero values of $A_0$ are observationally allowed.

\subsection{Dark energy }
\noindent
We briefly demonstrate here that it may be possible to use the LVK GW constraint to also place a bound on the cosmological constant. Recall the GW speed of propagation (\ref{cTUniversal}) that we found for nonzero $A_0$ in de Sitter space. 

Following the discussion in Sec.~\ref{sec:lvk-observations}, one can use this to constrain $m^2$ in terms of $\xi$ and $\Lambda$. By combining it with the propagation speed constraint from \cite{LIGOScientific:2017zic}, we find that $m^2$ must obey $m^2 \lesssim 2\Lambda\xi$.
Therefore, if $m^2$ satisfies this but violates the bound (\ref{eq:m2-ghost-bound}), we know we must be in the $A_0 = 0$ branch.

We can effectively think of (\ref{eq:m2-ghost-bound}) as a bound on the cosmological constant. As we saw earlier, the background evolution does not depend on the cosmological constant; this can be used to further constrain $\Lambda$, up to the other two parameters $m$ and $\xi$. In turn, we can think of this as a further hint to the cosmological constant problem, because we have to be in a branch where $\Lambda$ is small.

\subsection{Framework for matter coupling }\label{subsec:framework}

The non-minimally driven vector fields introduced in this work open a new framework that allows for non-trivial coupling to matter. One particular possibility that we have pointed out in Sec.~\ref{sec:matter} is to have a coupling of the form (\ref{s1caseCosmo}), with an arbitrary function $f$. We have previously explored the simplest case of this interaction, and found that if $m=0$ and $\Lambda=0$, then for constant values of $A_0$ one can also bring the equations of motion to the form that exactly matches the FLRW Universe. In the more general case, with time-evolving $A_0$, the above theory has a natural physical frame, known as the \textit{minimal frame}  \cite{Hell:2025lgn}, where matter would be minimally coupled both to the vector field and to gravity, reachable through a conformal transformation. 

In addition to this property for the $f(A_{\mu}A^{\mu})$ couplings, we would like to stress that our theory also allows for other possible couplings. Two interesting possibilities are the direct couplings between the energy-momentum tensor of matter $T^{\mu\nu}_{m}$ and the vector field, that in the simplest realization can be through its components directly, its trace, and even include further non-minimal coupling to gravity:
\begin{equation}
    \int d^4x\sqrt{-g} T^{\mu\nu}_{m}A_{\mu}A_{\nu},\qquad  \int d^4x\sqrt{-g} T_{m}A_{\mu}A^{\mu},\qquad  \int d^4x\sqrt{-g} RT_{m}A_{\mu}A^{\mu}, 
\end{equation}
Therefore, our vector field theory might open new exciting avenues with matter coupling that would nevertheless keep the same number of propagating degrees of freedom, up to the matter content.

\section{Discussion } \label{sec:discussion}

In this work, we have introduced a framework of specially constrained vector field theories, building on the recent work of constrained gravity, and we considered the simplest realizations of such theories in flat and cosmological spacetimes. We have found that such theories do not introduce any extra propagating modes, but they predict a modified GW dispersion relation. In fact, the GW propagation speed appears to be modified universally when $A_0$ is nonzero. 

Our analysis opens a new framework for vector field models through a non-minimally coupled matter window. In particular, for the cosmological solutions it avoids all issues that are otherwise found in theories driven by vector fields with non-minimal couplings, making this approach especially promising. While we have studied a few solutions and branches in this work, this can be generalized and we may have more solutions. For example, different black holes may be obtained for various values of $A_0$. This includes slowly rotating sources, for which the vector potential values might be different from GR. Importantly, we have found that our theory and GR both pass the solar system experiments; we recover the weak gravity limit in our theory for non-trivial $A_0$. Thus, both can be tested using e.g. the deflection of starlight or the precession of the perihelion of Mercury. One may also be able to study a stochastic GW background from our theory, and see how that compares with the sensitivities of current and upcoming pulsar timing experiments.

Furthermore, we have found that our model contains $\Lambda$CDM, but it also allows for other possibilities. Our theory could be used to constrain dark matter models -- either models with dark vector fields, or vector fields coupling to dark matter -- as well as other types of dark energy aside from the cosmological constant. For example, it may be possible to interpret our modified cosmology as giving rise to some form of evolving dark energy, and compute its equation of state. At the same time, our theory also allows for more general matter couplings, opening a new framework for vector fields. And, similarly to \cite{Hell:2025lgn}, in the case of a non-trivial $A_0$ branch, values of the cosmological constant do not affect any spacetime evolution. Therefore, this theory might be a further hint towards resolving the cosmological constant problem.

\acknowledgments

The authors would like to thank Robert Brandenberger, Elisa G. M. Ferreira and Misao Sasaki for useful discussions. A.H. thanks McGill University for hospitality, where part of this work was carried out, and Stephon Alexander, Robert Brandenberger, and Jia Liu for making this trip possible. The work of A.H. was in part supported by the KAKENHI No.25H00403, and is supported by the World Premier International Research Center Initiative (WPI), MEXT, Japan. T.D. is supported by a Trottier Space Institute fellowship and by funds from NSERC.

\begin{appendix}

\section{Perturbations around the de Sitter Universe}\label{appA}
In the following, we will analyze the branch with a non-vanishing value for the vector field, whose background evolution is given by (\ref{Hubble1}). Without any loss of generality, we can pick the positive branch solution $A_0=\sqrt{\frac{\Mpl^{2} \left(2 \Lambda \xi -m^{2}\right)}{2 \xi  \,m^{2}}, }$ and we perturb around the background values: \begin{equation}\label{dec}
    g_{\mu\nu}= g_{\mu\nu}^{(0)}+\delta  g_{\mu\nu}\qquad\text{and} \qquad A_{\mu}=A_{\mu}^{(0)}+\delta  A_{\mu},
\end{equation}
where $g_{\mu\nu}^{(0)}$ and $A_{\mu}^{(0)}$ are the background values. The perturbations of the metric and the vector can be decomposed in terms of the scalar, vector and tensor modes as: 
\begin{equation}\label{decomposition}
    \begin{split}
        \delta g_{00}&=-2\phi,\\
        \delta g_{0i}&=a(t)\left(S_i+B_{,i}\right),\\
        \delta g_{ij}&=a^2(t)\left(-2\psi \delta_{ij}+2E_{,ij}+F_{i,j}+F_{j,i}+h_{ij}^T\right),\\
         \delta A_{0}&=A,\\
         \qquad \delta A_i&=a(t)A_i^T+\chi_{,i}, 
    \end{split}
\end{equation}
where the comma denotes the derivative with respect to the spatial component. The vector and tensor modes satisfy: 
\begin{equation}
    S_{i,i}=0,\qquad F_{i,i}=0,\qquad  A_{i,i}^T=0,\qquad h_{ij,j}^T=0\qquad \text{and}\qquad h_{ii}^T=0. 
\end{equation}

The quadratic actions for scalar, vector and tensor perturbations obtained by expanding the action to second order decouple, and we can study them separately. 

Let us first consider the scalar modes. Integrating the quadratic action by parts and analyzing the kinetic matrix, we can notice that all scalars apart from $\psi$ are not propagating. We then find the constraint for $\phi$ by varying the action with respect to it. By solving the constraint for the same field and substituting back into the action, we find an expression that is a function of only $\chi,$ $A$ and $\psi$. However, both $\chi$ and $A$ again do not propagate. We then find the constraint for $\chi$, solve it with respect to $\chi$, and substitute back to the action. This cancels the kinetic term for $\psi$, making it also a non-propagating field. Finally, the constraint for $A$ yields $\psi=0$, which then leads us to $A=0$. Therefore, this theory for the homogeneous and isotropic background propagates no scalar degrees of freedom. 

Similarly, one can show that the vector modes do not propagate as well. Both vector perturbations arising from the metric perturbations and the vector field lack a kinetic term. Thus, by resolving one of the constraints, one will be led to an additional set of constraints, and finally find that the action vanishes.

\section{Departing from the Einstein tensor coupling}\label{gendof}
In the main analysis of this work, we have considered the coupling of the constrained vector field to the Einstein tensor. In this Appendix, we will outline a simple example of cosmological expansion showing that for more general couplings, the theory will propagate an additional scalar mode, and therefore our choice of coupling in Sec.~\ref{sec:theory} is special. For this, let us consider the action: 
\begin{equation}\label{pureVFgen}
    S=\int d^4x\sqrt{-g}\left[\frac{\Mpl^2}{2}\left(R-2\Lambda\right)-m^2A_{\mu}A^{\mu}+\frac{1}{2}\xi_1 RA_{\mu}A^{\mu}-\xi_2 R^{\mu\nu}A_{\mu}A_{\nu}\right], 
\end{equation}
where the couplings $\xi_1$ and $\xi_2$ are now arbitrary. For a background given by the ansatz (\ref{ansatzdS}), we find the following equations:
\begin{equation}
   0= A_{0} \left(-6 a \left(\xi_{1}-\xi_{2}\right) \ddot{a}+m^{2} a^{2}-6 \xi_{1} \dot{a}^{2}\right)
\end{equation}
by varying the action (\ref{pureVFgen}) with respect to the $A_0$ component,
\begin{equation}\label{constraintgenC}
    0=6 \left(\left(-\xi_{1}+2 \xi_{2}\right) A_{0}^{2}+\Mpl^{2}\right) \dot{a}^{2}-12 a \dot{A}_0 A_{0} \left(\xi_{1}-\xi_{2}\right) \dot{a}+a^{2} \left(A_{0}^{2} m^{2}-2 \Lambda \,\Mpl^{2}\right)
\end{equation}
by varying (\ref{pureVFgen}) with respect to the lapse, and
\begin{equation}
    \begin{split}
        0&=\left(\left(-2 \xi_{1}+4 \xi_{2}\right) A_{0}^{2}+2 \Mpl^{2}\right) \dot{a}^{2}-8 a \dot{A}_0 A_{0} \left(\xi_{1}-2 \xi_{2}\right) \dot{a}-4a^2 \left(\xi_{1}-\xi_{2}\right) \dot{A}_0^{2}\\&-4 \left(\left(\left(\xi_{1}-2 \xi_{2}\right) A_{0}^{2}-\Mpl^{2}\right) \ddot{a}+a \left(A_{0} \left(\xi_{1}-\xi_{2}\right) \ddot{A}_0-\frac{A_{0}^{2} m^{2}}{4}+\frac{\Lambda \,\Mpl^{2}}{2}\right)\right) a
    \end{split}
\end{equation}
by varying (\ref{pureVFgen}) with respect to the scale factor. We can see right away from (\ref{constraintgenC}) that for $A_0\neq 0$, the theory will admit solutions more general than the de Sitter Universe, and reduce to the original case for $\xi_1=\xi_2$. In this section however, we will focus mainly on the perturbations, and not the solutions themselves. To ensure that the background equations of motion are always satisfied, we will express the above three equations in terms of $\Lambda,$ $\ddot{a}$ and $\ddot{A}_0$, and assume that they are always satisfied at the level of action. 

In what follows, we will outline the procedure to obtain the propagating scalar, although we will omit the actual expressions up to the most important one. Let us next perturb the metric and the vector field according to (\ref{dec}) and (\ref{decomposition}), and focus on the scalar modes, expanding the action up to second order in the perturbations. First, without loss of generality, we can fix the longitudinal gauge $E=0$ and $B=0$.  Integrating by parts and substituting $A=A_2+\phi A_0$, we find that $\phi$ stops propagating in addition to the $\chi$ mode. By varying the action \eqref{pureVFgen} with respect to $\phi$, solving the constraint and substituting it back to the action, we find the expression in terms of the fields $A$, $\psi$ and $\chi$. Next, after integrating by parts, we substitute $\psi=\psi_2+\frac{\dot{a}}{a\dot{A}_0}A_2$, which makes $A_2$ and $\chi$ non-propagating fields. We find the constraint for $\chi$, solve it, and substitute back to the action, which then becomes a function of only two fields: $A_2$ and $\psi_2$. Integrating by parts, we find that $A_2$ does not propagate. We again find its constraint, solve it, and substitute back to the action. This results in an action that contains only $\psi_2$, with a kinetic term containing $\dot{\psi}_2^2$ which is in the momentum space given by: 
\begin{equation}
       \frac{12 C a^{5} \dot{A}_0^{2} A_{0}^{2} \left(\xi_{1}-\xi_{2}\right)^{2}}{\left(\left(\xi_{1}-2 \xi_{2}\right) A_{0}^{2} \dot{a}+\left(\xi_{1}-\xi_{2}\right) \dot{A}_0 A_{0} a -\Mpl^{2} \dot{a}\right)^{2} \left(6 \left(-2 \xi_{1}+\xi_{2}\right) \dot{a}^{2}+m^{2} a^{2}\right)},
\end{equation}
where 
\begin{equation}
  \begin{split}
       C&= -\left(-3 \left(\xi_{1}-\frac{\xi_{2}}{2}\right) \left(\xi_{1}-2 \xi_{2}\right) \dot{a}^{2}+\frac{\left(\xi_{1}-2 \xi_{2}\right) m^{2} a^{2}}{4}+\left(\xi_{1}-\xi_{2}\right) k^{2} \xi_{2}\right) A_{0}^{2}\\&+\frac{\Mpl^{2} \left(6 \left(-2 \xi_{1}+\xi_{2}\right) \dot{a}^{2}+m^{2} a^{2}\right)}{4}.
  \end{split}
\end{equation}
Notably, the above kinetic term vanishes for $\xi_1=\xi_2$, which corresponds to the Einstein tensor coupling we have previously considered. We can therefore see that unless this type of coupling is chosen, the theory will propagate a scalar mode around a cosmological background.

\end{appendix}
\bibliography{bibliography}

\end{document}